\documentstyle[a4]{article}

\begin{document}

\begin{center}
{\LARGE{\bf{Magnetic Field Induced Two-Body Phenomena in Atoms}}}
\end{center}

\vspace{0.3cm}
\begin{center}
{\large{\bf{Peter Schmelcher and Lorenz S. Cederbaum}}}
\end{center}
\begin{center}
Theoretische Chemie, Universit\"at Heidelberg, 
Im Neuenheimer Feld 229, 69120 Heidelberg, 
Federal Republic of Germany
\end{center}

\begin{abstract}
\noindent
We report on a number of recently discovered phenomena which arise
due to the interaction of the collective (CM) and internal motion of
atoms moving in magnetic fields. For neutral atoms
the properties of the so-called giant dipole states are discussed
and their experimental preparation is outlined.  
For ions moving in magnetic fields the energy flow between
the collective and electronic motions is studied. For low CM velocities
this results in a crossover from localization to delocalization with
respect to the spreading of the mixing of the CM and internal motion
in quantum number space. Large CM velocities induce for both the
classical as well as quantum ion the self-ionization process. 
Future perspectives for both multi-electron atoms as well as molecules
are drawn.

\end{abstract}

\vspace{0.3cm}

\section{Introduction}
External magnetic fields are well-known to have a strong impact on the
properties of particle systems. A number of
intriguing phenomena in different areas of physics like, for example,
the quantum Hall effect in solid state physics or the interplay of
regularity and chaos in few-body atomic systems \cite{Fri89} have their origin
in the combination of the magnetic and Coulomb forces.
By nature the Coulomb potential together with the paramagnetic and diamagnetic 
interaction represents a nonseparable and nonlinear problem already on the
one-particle level. As a consequence both the dynamics as well as the
electronic structure of atoms or molecules \cite{Schm98a} are subject to severe changes
in the presence of a strong magnetic field. 

In the present article we review on selected recent developments
for Rydberg atoms in strong laboratory magnetic fields
and outline future perspectives. We also provide concrete suggestions
how to experimentally access and observe the corresponding atomic states and processes.
Of course, similar phenomena are to be expected also for astrophysical field strengths
but the focus in this work is on highly excited atoms exposed to laboratory field strengths.
Our central issue are phenomena and effects arising due to the
nonseparability of the center of mass and internal motion for two-body systems
in external magnetic fields and/or crossed electric and magnetic fields \cite{Schm97a}.
Due to the smallness of the ratio of the electron to nuclear mass one might be, at 
first glance, tempted to believe that the effects of the coupling of the collective and
internal motion provide only tiny corrections to the overall dynamics. This is 
however in general not correct and indeed we will provide a number of different physical 
situation in which even the correct qualitative behaviour cannot be obtained
without including the interaction of the center of mass and relative motion.
Prominent effects due to this interaction are the chaotic classical diffusion of the
center of mass \cite{Schm92a,Schm92b,Schm92c,Fri98}, the existence of giant dipole states for neutral atoms
and dynamical phenomena like the self-ionization process for rapidly moving
ions in strong magnetic fields. As we shall see in the following the ongoing research
in this area reveals more and more of the beautiful phenomena created by the
competition of magnetic and Coulomb forces.

In detail we proceed as follows. Section I gives first a brief account of the separation
of the CM motion for neutral particle systems in magnetic fields and subsequently
investigates a class of decentred magnetically dressed Rydberg states
possessing a huge electric
dipole moment. A possible experimental set-up for the preparation of these states as well as
applications to the positronium atom are discussed. Section II reviews on recent developments
for ionic systems like the classical self-ionization process for rapidly moving ions in magnetic fields.
A number of most recently discovered quantum properties of ions in fields are outlined and clearly
indicate that quantization introduces a number of new features.

\section{Neutral atoms in magnetic fields}
\subsection{Giant dipole states in crossed fields}

In the absence of an external magnetic field the CM and electronic motions of an atom
separate due to the conservation of the total canonical momentum $\bf{P}$. 
$\bf{P}$ is the generator for coordinate space translations which represent a symmetry
for any system of particles interacting through a potential which depends only on the
distances of the particles. 
In a homogeneous magnetic field this symmetry is lost since the Hamiltonian depends
explicitly on the vector potential associated with the external field. Nevertheless,
there exists a so-called phase space translation group which represents a symmetry
in the presence of the external field. This group is generated by the so-called pseudomomentum.
Historically the pseudomomentum was implicitly used by Lamb \cite{Lam52} in order to perform
a so-called pseudoseparation of the CM motion for the hydrogen atom in a homogeneous
magnetic field. In the late seventies a more profound mathematical treatment of the
pseudoseparation for two-body systems was achieved \cite{Avr78}.
In the eighties and nineties the pseudoseparation
for many-body systems has been reviewed (see ref.\cite{Joh83} and references therein)
and it was adapted to the needs of molecular physics \cite{Schm94a}.

In order to perform the pseudoseparation transformation as mentioned above
one starts with the Hamiltonian of the atom in the laboratory frame which
reads for a two-body system
\begin{equation}
 {\cal H}_{L} = \sum\limits_{i=1}^{2}\lbrack
 \frac{1}{2m_i}({\bf{p}}_i - e_{i} {\bf{A}}_i )^2 -
 e_{i}{\bf{E}}{\bf{r}}_i\rbrack
 + V(|{\bf{r}}_1-{\bf{r}}_2|)
\end{equation}
where $e_{i},m_{i},{\bf A}_{i}$ and ${\bf E}$ denote the charges, masses,
vector potential and electric field vector, respectively.
$\lbrace{\bf r}_{i},{\bf p}_{i}\rbrace$ are the Cartesian coordinates
and momenta in the laboratory coordinate system.
Subsequently one chooses a certain gauge for the vector potential ${\bf A}_{i}$
belonging to the magnetic field ${\bf B}$. Most commonly this is done
by introducing the symmetric gauge ${\bf A}_{i} = \frac{1}{2} {\bf B} \times {\bf r}_i$.
As a next step a coordinate transformation from laboratory to relative and center of 
mass coordinates follows. Finally the {\it pseudomomentum} is introduced as a {\it canonical 
momentum of the center of mass coordinate} by performing a corresponding transformation in momentum space
which gives the final transformed Hamiltonian.
Choosing, according to the above, a specific gauge in the laboratory Hamiltonian
has, however, a serious drawback: It is not possible to
discern between gauge dependent and gauge invariant terms in the transformed
Hamiltonian. In order to extract the physics, which is inherently gauge-independent,
from the transformed Hamiltonian it is a central issue to identify gauge invariant parts of it.
The necessity of a gauge-invariant approach has been realized
in the nineties by different authors \cite{Dzy92,Bay92,Dip94} which have chosen different
approaches in order to obtain a gauge-independent formalism and results for the problem
of two interacting particles. Most importantly it was finally shown \cite{Dip94,Schm93a} that
there exists a potential picture (see below) due to a number of gauge-invariant terms
emerging from the generalized pseudoseparation.

Let us consider the final Hamiltonian resulting from the afore-mentioned
gauge independent pseudoseparation \cite{Dip94}
\begin{equation}
{\cal{H}}={\cal{T}}+{\cal{V}}
\end{equation}
where
\begin{equation}
{\cal{T}}=\frac{1}{2\mu}({\bf{p}}-q{\bf{A}}({\bf{r}}))^2
\end{equation}
and
\begin{equation}
{\cal{V}}=\frac{1}{2M}({\bf{K}}-e{\bf{B}}\times{\bf{r}})^2+V(r)
+\frac{M}{2}{\bf{v}}_{D}^2+{\bf{K}}{\bf{v}}_{D}
\end{equation}
with the charge $q=\frac{e\mu}{\hat{\mu}}$ where
$\mu=\frac{mM_{0}}{M}$ and ${\hat{\mu}}=\frac{mM_{0}}{M_{0}-m}$
are different reduced masses. ${\bf{K}}$ is now the constant vector of
the pseudomomentum which contains a term linear in the external electric
field. $\lbrace{\bf{r}},{\bf{p}}\rbrace$ denote
the canonical pair of variables for the internal relative motion.
${\bf{v}}_{D}= \frac{({\bf E}\times{\bf B})}{B^2}$ is the drift velocity of free charged
particles in crossed external fields. The latter is independent of the charges and
masses of the particles.
The Hamiltonian (1) is the sum of two terms: The kinetic energy ${\cal{T}}$
of the relative motion and the potential ${\cal{V}}$.

The explicit form of the kinetic energy ${\cal{T}}$ depends on the chosen
gauge via the vector potential
${\bf{A}}$. The important novelty with respect to our Hamiltonian
${\cal{H}}$ is the appearance of the potential term ${\cal{V}}$. Apart from the
Coulomb potential $V$ and the constant term
$\frac{M}{2}{\bf{v}}_{D}^2+{\bf{K}}{\bf{v}}_{D}$
${\cal{V}}$ contains an additional potential term 
\begin{equation}
V_{o}=\frac{1}{2M}({\bf{K}}-e{\bf{B}}\times{\bf{r}})^2
\end{equation}
The latter is gauge independent, 
i.e. does not contain the vector potential, and, therefore, fully deserves
the interpretation of an additional potential for the internal
motion with the kinetic energy (2). Apart from the constant
$\frac{{\bf{K}}^2}{2M}$ the potential $V_{o}$ contains two
coordinate dependent parts. The term linear in the coordinates
$-\frac{e}{M}({\bf{K}}\times{\bf{B}}){\bf{r}}$ consists of two
Stark terms: one which is due to the external electric field
${\bf{E}}$ and a second one which is
a motional Stark term with an induced constant electric field
$\frac{1}{M}(({\bf{K}}+M{\bf{v}}_D)\times{\bf{B}})$.
The latter electric field is always oriented perpendicular to the magnetic one
and arises due to the collective motion of the atom through the
homogeneous magnetic field. Besides the linear terms there exists a
quadratic, i.e. diamagnetic, term $\frac{e^2}{2M}({\bf{B}}\times{\bf{r}})^2$
in the potential $V_{o}$ which is of great importance for the main
subject of the present section, i.e. the existence of giant dipole 
states of two-body system in external fields.

In the following we discuss the qualitative properties of
the potential ${\cal{V}}$. Figure 1 shows for the choice ${\bf{B}}=(0,0,B)$,
${\bf{K}}=(0,K,0)$ and a vanishing external electric field a two-dimensional
intersection of the potential ${\cal{V}}$ for $z=0$.
In the neighbourhood of the coordinate origin the Coulomb potential dominates.  
This regime corresponds in figure 1 to the tube around the origin
of the coordinate plane $(x,y)$.
With increasing distance from the origin along the $x-$axis
($y=0$) the Stark term $(\frac{-e}{M}
)BKx$ increases and becomes eventually comparable with the Coulomb
potential. This means that we encounter an approximately linear raise 
of the potential for positive values of the $x-$coordinate.
For negative values of the $x-$coordinate eventually a
saddle point emerges. In this coordinate regime the diamagnetic term of eq.(4) 
represents a small correction. For even larger distances 
the Coulomb potential becomes small and the shape of the
potential  ${\cal{V}}$ is more and more determined by the diamagnetic
term $(\frac{e^2}{2M}){B^2}{(x^2+y^2)}$. Due to the
competition of the Stark and diamagnetic terms our potential ${\cal{V}}$
develops an outer minimum and a corresponding potential well.
The existence of both the saddle point
and the outer minimum/well depends, of course, on the values of the magnetic
field strength and the motional/external electric 
field strength. For a derivation and discussion of the 
corresponding conditions we refer the reader
to the literature \cite{Dip94,Schm93a}.
We emphasize that the potential $V_{o}$ is inseparably
connected with the finite nuclear mass. Assuming an infinite nuclear
mass simply yields $V_{o}\equiv 0$ and the described
features of the total potential ${\cal{V}}$ disappear. 

The above-discussed properties of our potential ${\cal{V}}$ have
important implications on the dynamical behaviour of the atom.
First of all, we observe that the ionization of the atom can take place
only in the direction of the magnetic field axis: In the direction
perpendicular to the magnetic field vector the diamagnetic term
$\frac{e^2}{2M}({\bf{B}}\times{\bf{r}})^2$ is dominating for large
distances $\rho=(x^2+y^2)^{1/2}$ and causes a confining behaviour.
The second important observation is the fact that
the existence of the outer well leads to new weakly bound states in this well.
Let us provide some main characteristics of these states. 
In refs.\cite{Dip94,Schm93a} an explicit approximation formula has been given for the 
position of the outer minimum: $x_0\simeq -\frac{K}{B}+\frac{KM}{K^3-2MB},y=z=0$.
Hence, for a laboratory field strength $B\sim 2.35~Tesla$ and 
a motional/external electric field of the order of $E \sim 2.8 \times
10^{3} \frac{V}{m}$ the minimum is located at a distance of
about $5.3 \times 10^{-6} m$ from the Coulomb singularity.
For states in the outer well the electron and proton are therefore
separated about 100,000 times as much as they are in the ground state
of the hydrogen atom without external fields, i.e. we encounter
a strongly delocalized atom of mesoscopic size.
Since the well exists only on the half-hyperplane with a negative
Stark term (see fig.1)  these states possess a huge permanent electric dipole
moment of the atom. This is in contrast to the well-known Rydberg states
in a pure magnetic field which do not exhibit a permanent dipole moment.
For energies close to the minimum the outer well is approximately an
anisotropic harmonic potential. Low-lying quantum states can 
therefore be described by an anharmonic oscillator in a magnetic field
\cite{Dip94}. The field-dependent kinetic energy (see eq.(2)) hereby
determines the extension of the wave function in the plane perpendicular
to the magnetic field. The deviation of the exact energies
from those of the harmonic approximation changes, as expected, significantly
with increasing degree of excitation. It grows stepwise and a major
contribution to the anharmonicity comes from the quantum number $n_z$
i.e. the excitation in the direction parallel to the magnetic field.
This can also be seen in perturbation theory for higher terms of the expansion
of the Coulomb potential where the major contributions to the energy
corrections are due to those terms containing high powers of $z$.
For the computational techniques to obtain energies and wavefunctions
for a large number of states in the outer well
as well as a detailed discussion of their properties
we refer the reader to ref.\cite{Dip94}.

\subsection{Experimental preparation} 

At this point it is natural to pose the question how one can experimentally prepare hydrogen
atoms in crossed fields in their giant dipole states. This question has been 
investigated and discussed in detail in ref.\cite{Ave99}. We provide here the
key ideas of the approach developed in this work. The important parameter
which controls the formation of the outer well and is at disposal to the experimentalist
is the external electric field strength. The preparation scheme consists
of a sequence of steps which correspond to different electric field configurations.
The first step excites the atom in the presence of a magnetic field but no
electric field from the ground state with a laser pulse to a highly excited
Rydberg state. In the second step an electric field is switched on within a 
time period of a few $ns$ to a value $E=E_c$ which corresponds to the existence
of a shallow outer well. Subsequently, during the third step the electric
field is kept constant for a time period $\Delta t$.
The Rydberg states prepared in step one are localized in the Coulomb well.
After the switching of the electric field to $E_c$ (step two) their energy
is above the saddle point. Keeping the electric field constant during the third time step
has the reason to achieve a significant spreading of the prepared
state over the shallow outer well. Thereafter, in a third step, the electric
field is switched to its final value $E_f$ which corresponds to a deep outer
well. This last step of the preparation procedure captures the wave packet 
in the outer well and a significant portion of the hydrogen atoms
therefore ends up in low-energy states of the outer well. The second switching
of the electric field to its final value $E_f$ is much slower than the first
fast switching to the value $E_c$. The reason herefore is the broadening of the
final energy distribution during the trapping of the hydrogen atoms in the outer well.
An adiabatic second switching is therefore obligatory in order to end up with
a narrow final energy distribution. Figure 2 shows the final energy distribution
in the outer well resulting from calculations for an ensemble of  classical
trajectories which simulate the behaviour of the highly excited Rydberg states
during the above-discussed steps of preparation. It can clearly be seen that
the maximum of the energy distribution is below the ionization threshold
(vertical dashed line in figure 2) and therefore a significant part of the
prepared outer well states is strictly bound. For further details of the
experimental setup and preparation like, for example,
the specific switching procedures of the electric field or the influence of electric stray 
fields on the preparation scheme we refer the reader to ref.\cite{Ave99}.

Let us provide some remarks with respect to the
experimental detection of the giant dipole states of the hydrogen atom
in crossed electric and magnetic fields. Direct state-to-state
transitions for bound states in the outer well should 
be observable in the radio-frequency regime. 
The energy gap between the ground and first excited state
corresponds, for the above-given field strengths,
to a frequency of the order of magnitude of a few tens of MHz.
Alternatively the large electric dipole moment suggests itself
for detection which could be achieved through
deflection of the atoms by a slightly inhomogeneous electric field \cite{Fau87}.
Even though the binding energies of the states in the outer well are relatively
small they should be stable as long as collisional interaction is 
prevented.

We have discussed the existence and properties as well as the experimental preparation and
detection of giant dipole states for the hydrogen atom in crossed fields.
The electric field can hereby be either an external one or due to
the collective motion of the hydrogen atom through the magnetic field. 
For these states the magnetic interaction is dominating the Coulomb attraction
in the plane perpendicular to the magnetic field whereas parallel to the magnetic
field we exclusively encounter Coulomb forces. The electric field causes the
decentred character of these states \cite{Bez94,Pot94,Bez95}. Of course this kind of
states does not only exist for one-electron atoms but should occur also for
more-electron systems. Indeed it is a challenging question
to ask for the existence and properties of decentred multielectron 
atoms in crossed external fields. Moreover one can imagine giant dipole
molecules which, due to the presence of several heavy particles,
might possess completely different features compared to atoms.

\subsection{Application to Positronium}

Besides the above-drawn general perspective there exists an intriguing application
of the above results to exotic two-body systems, namely the positronium atom
\cite{Ack97,She98}. Since the distance of the minimum of the outer well from
the Coulomb singularity is approximately proportional to the total mass of the two-body system
$(\propto \frac{EM}{B^2})$ we expect
that the giant dipole states for positronium are significantly less extended
than those of hydrogen. At the same time the critical electric
field strength necessary for the existence of the outer well scales with $M^{-\frac{2}{3}}$.
As a consequence the typical reduction of size for fixed field strength
is of the order of $10$, i.e. for $B\sim 2.35~T$ 
the extension of the Rydberg state is several thousand Bohr radii.
%Considering energies corresponding to low-lying states in the outer well
%it was shown \cite{Ack97,She98} that the bound state spectrum corresponds to
%that of two isolated potential wells: a set of Coulomb well states localized
%at the origin and a set of outer well states centered around the outer
%minimum.
The giant dipole states located in the outer well and the traditional Rydberg states
located in the Coulomb well are separated by a wide and high potential barrier.
For the positronium atom this has important
consequences: the potential barrier prevents the particles from contact
and therefore decreases the annihilation rate by many orders of magnitude.
Indeed, for typical laboratory field strengths the lifetime can become many years and low-lying
outer well states of positronium can, for all practical purposes, be considered
as stable \cite{Ack97,She98}. Crossed fields offer therefore a unique opportunity for the
stabilization of matter-antimatter two-body systems. For a detailed investigation
of the positronium atom including dipole transition rates, tunneling probabilities
as well as spectra and wavefunctions resulting from both perturbation theoretical
and finite element calculations we refer the reader to refs.\cite{Ack97,She98}.
We conclude with an important result of ref.\cite{She98} which should be of
relevance to astrophysical situations:
For sufficiently strong fields the energetically lowest decentred outer well state
becomes the global ground state of the atom. This statement holds for both the
hydrogen atom as well as the Positronium atom. As a consequence the ground state
of isolated positronium in strong crossed external fields is prevented
from annihilation and represents a long-lived state.

\section{Atomic ions in magnetic fields}
\subsection{Basic properties}

For a charged atom in a magnetic field the interaction of its collective motion 
with the electronic motion is more
intricate than for neutral species. From a physical point of view it is evident
that the crudest picture for the CM motion describes the ion as
an entity in the magnetic field with a charge and mass identical to the
net charge and total mass of the ion. For neutral systems the net charge of the
system is zero and the crudest picture for the behaviour of the CM 
is a free straightlined motion through the magnetic field.
However, as we shall see below, we encounter coupling terms for the ions 
CM and electronic motions which can mix them up heavily thereby
causing a number of interesting energy transfer processes between the degrees of freedom. 

From a formal point of view the maximum number of commuting constants of motion
is the same for neutral as well as charged systems. For neutral species, however,
one can choose the three components of the total pseudomomentum $\bf K$
which are exclusively
associated with the CM motion of the system. As a consequence the CM coordinates
are cyclic and can be completely eliminated by the so-called pseudoseparation
from the Hamiltonian (see section 2.1). For charged systems the only maximal
set of commuting constants of motion is $(\bf{K}_{\perp}^2,\cal{L}_{
\parallel},{\bf{K}}_{\parallel})$
where ${\bf{K}_{\perp}}$ is the component of the pseudomomentum perpendicular
to the magnetic field and $\cal{L}_{\parallel}$
is the component of the total angular momentum parallel to the field.
This set of quantities is not exclusively associated with the CM motion but involves
both the CM and internal degrees of freedom. Unlike the neutral system the
CM coordinates cannot completely be eliminated from the Hamiltonian. 
To simplify the Hamiltonian one therefore uses a transformation \cite{Schm91a} which
possesses a certain analogy to the transformations in the neutral case
thereby arriving at the following final form
\begin{equation}
{\cal H} = {\cal H}_1 + {\cal H}_2 + {\cal H}_3
\end{equation}
where
\begin{equation}
{\cal H}_1 =  \frac{1}{2M}\left({\bf{P}}-\frac{Q}{2}{\bf{B}}\times
{\bf{R}}\right)^2
\end{equation}
\begin{equation}
{\cal H}_2 = e\frac{\alpha}{M}\left({\bf{B}}\times\left({\bf{P}}
-\frac{Q}{2}{\bf{B}}\times
{\bf{R}}\right)\right){\bf{r}}
\end{equation}
\begin{eqnarray}
{\cal H}_3 & = & \frac{1}{2m}\left({\bf{p}}-\frac{e}{2}{\bf{B}}\times
{\bf{r}}+\frac{Q}{2}\frac{m^2}{M^2}{\bf{B}}\times{\bf{r}}\right)^2 \\ \nonumber
& & +\frac{1}{2M_0}\left({\bf{p}}+\left(\frac{e}{2}-\frac{Q}{2M}\frac
{m}{M}\left(M+M_0\right)\right){\bf{B}}\times
{\bf{r}}\right)^2 +V
\end{eqnarray}
where $m,M_0$ and M are the electron, nuclear and total
mass, respectively. $\alpha=(M_0+Zm)/M$ and V is the Coulomb potential.
For the vector potential ${\bf{A}}$ we have adopted the symmetric gauge.
The magnetic field vector $\bf{B}$ again is assumed to point along the z-axis.
$({\bf{R}},{\bf{P}})$ and $({\bf{r}},{\bf{p}})$ are the canonical pairs
of variables for the CM and relative motion, respectively. 

The Hamiltonian ${\cal H}$ consists of three parts which
introduce different types of interaction. The part
${\cal H}_1$ in eq.(6), which involves solely the CM degrees of freedom,
describes the cyclotron motion of a free pseudoparticle with mass M
and charge Q in a homogeneous magnetic field (see comments
at the beginning of this subsection).  
This zeroth order picture is, in general, not sufficient to describe
the CM motion of the ion. In fact the behaviour of the CM can deviate strongly from the
motion given by the Hamiltonian ${\cal H}_1$ and exhibits a variety of
different phenomena depending on the parameter values
(energy, field strength, CM velocity)  \cite{Schm91a,Schm95a,Schm95b,Lei98,Mel99a,Mel99b}.
The origin of this rich dynamics lies in particular
in the Hamiltonian ${\cal H}_2$ in eq.(7) which describes the coupling
of the CM and electronic degrees of freedom. It
represents a motional Stark term with a rapidly oscillating electric field
which is determined by the dynamics of the system. Because of this
"dynamical" electric field the collective and internal motion will,
in general, mix up heavily. ${\cal H}_3$ in eq.(8)
contains only the electronic degrees of freedom and describes, to zeroth
order, the relative motion of the electron with respect to the nucleus.

\subsection{Mixing and localization properties for the collective and electronic
motions}

The Hamiltonian ${\cal H}_2$ in eq.(7) destroys the picture of a
decoupled CM and electronic motion and causes an interaction of these motions
whose strength depends on several parameters (CM and internal energy, field
strength). In the following we outline both classical and quantum
effects and phenomena due to this interaction 
\cite{Schm91a,Schm95a,Schm95b,Lei98,Mel99a,Mel99b}. We hereby pursue the track
of an increasing strength of the coupling of the collective and internal motion
which can in particular be achieved by increasing the energy of the CM motion.

In order to investigate the significance and effects of the coupling Hamiltonian
${\cal H}_2$ one can use the following method for formally solving the total
Schr\"{o}dinger equation ${\cal{H}}\Psi= E\Psi$. The most natural way is to
expand the total wave function $\Psi$ in a series of products
\begin{equation}
\Psi({\bf{R}};{\bf{r}})=\sum\limits_{p,q}{}c_{pq}\Phi_{p}^{L}(
{\bf{R}})\psi_{q}({\bf{r}})
\end{equation}
where $c_{pq}$ are the coefficients of the product expansion.
The functions $\lbrace\Phi_{p}^{L}\rbrace$ obey the Schr\"{o}dinger
equation ${\cal{H}}_{1}\Phi_{p}^{L}=E_{p}^{L}\Phi_{p}^{L}$ for a free
particle with charge $Q$ and mass $M$ in a homogeneous magnetic field,
i.e. they are the corresponding Landau orbitals.
One proper choice for the functions $\lbrace\psi_{q}\rbrace$ in the expansion (9)
(see also below) are the eigenfunctions of the electronic Hamiltonian, i.e. ${\cal{H}}_{3}
\psi_{q}=E_{q}^{I}\psi_{q}$ (q stands collectively for all electronic
quantum numbers). If we insert the product expansion (9) for the
total wave function $\Psi$ in the total Schr\"{o}dinger equation
and project on a simple product $\Phi_{p'}^{L}\psi_{q'}$ we arrive
at the following set of coupled equations for the coefficients
$\lbrace c_{pq}\rbrace$:
\begin{equation}
({\underline{\cal{H}}}_{2}+{\underline{E}}^{L}+{\underline{E}}^{I}
){\bf{c}}=E{\bf{c}}
\end{equation}
where ${\bf{c}}$ is the column vector with components $\lbrace c_{pq}
\rbrace$. ${\underline{E}}^{L}$ and ${\underline{E}}^{I}$ are the
diagonal matrices which contain the Landau energies $\lbrace E_{p}
^{L} \rbrace$ and internal energies $\lbrace E_{q}^{I} \rbrace$,
respectively. The original problem of the investigation of the
significance and effects of the couplings among the CM and electronic
degrees of freedom is now reduced to that of the solution of the
eigenvalue problem (10). ${\underline{\cal{H}}}_{2}$ contains the matrix
elements of the coupling Hamiltonian ${\cal{H}}_{2}$.
These elements can, via certain commutation relations, be transformed
into pure dipole transition matrix elements of the CM as well
as internal degrees of freedom (see ref.\cite{Schm91a} for details)
\begin{equation}
{\underline{\cal{H}}}_{2}=-ie\alpha (E_{p'}^{L}-E_{p}^{L})
\cdot \langle \Phi_{p'}^{L}
\vert {\bf{R}} \vert \Phi_{p}^{L} \rangle \cdot
\lbrack {\bf{B}}\times \langle \psi_{q'}\vert {\bf{r}} \vert
\psi_{q} \rangle \rbrack
\end{equation}
Obviously the coupling matrix elements $({\underline{\cal{H}}}
_{2})_{\lbrace p'q'
\rbrace \lbrace pq \rbrace }$
vanish if the two Landau states $\Phi_{p'}^{L}$ and $\Phi_{p}^{L}$
possess the same energy $E_{p'}^{L}=E_{p}^{L}$. Transitions, therefore,
do occur only for total states which involve 
different states of the collective motion with respect to their energy.

The expansion (9) with $\lbrace\psi_{q}\rbrace$ being the eigenfunctions
of the electronic Hamiltonian ${\cal{H}}_{3}$ 
is particularly adequate if the coupling is sufficiently
weak which means that the series contains only a few dominant terms.
Otherwise, i.e. in the case that many terms contribute significantly to the sum,
the mixing of the CM and electronic motion is strong and may require for its
proper and efficient description either a different choice for the functions
$\lbrace\psi_{q}\rbrace$ or an alternative approach (see below) to the expansion (9). 
%Additionally the evaluation of the electronic dipole matrix elements
%in eq.(11) can become very tedious if not impossible for highly excited
%Rydberg states (for an alternative see the semiclassical evaluation of coupling matrix elements
%in ref.\cite{Lei98} and below).
Let us evaluate the importance of the coupling
terms for different physical situations at laboratory field strengths.
We hereby concentrate on the $He^{+}$-ion moving in a magnetic field.
The simplest way to treat the couplings is to assume that the electronic wave functions are
to zeroth order well-described by the field-free hydrogenic wave functions
and to take into account the electronic diamagnetic interaction via
perturbation theory. This description is appropriate for laboratory field strengths
for not too high electronic excitations, i.e. for a field strength of a 
few Tesla typically up to $n=10-20$.  The relevant indicator for the mixing of CM and
electronic motions is the quotient of the coupling
$\kappa$ and the energy spacing $\Delta$ of the corresponding diagonal
matrix elements in $({\underline{E}}^{L}+{\underline{E}}^{I})$.
For electronic states belonging to the same principal quantum number $n$
(the latter being not too large) i.e. for the case of dominant intramanifold
coupling, we obtain \cite{Schm91a}
\begin{equation}
(\kappa/\Delta)\propto\frac{\sqrt{NB}}{M}n^{2}
\end{equation}
where $N$ is the principal quantum number of the Landau orbitals.
For $n=10$ and a typical laboratory field strength $B=2.35 T$ 
$N$ has to be of the order of magnitude $N\approx 10^{8}$
to make the coupling $\kappa$ as large as the energy spacing $\Delta$.
The corresponding energy of the CM motion is some $10 eV$. 
For the case of intermanifold coupling, i.e. the coupling of states
belonging to different $n$-manifolds we obtain
\begin{equation}
(\kappa/\Delta)\propto\frac{\sqrt{NB}}{M}Bn^{5}
\end{equation}
Choosing $B=10 T$ and $n=10$ the requirement that
$\kappa$ should be of the order of magnitude
of $\Delta$ yields $N\approx 10^{10}$, i.e. a CM energy of the ion of a
few keV. This means that at these energies the couplings become
not only dominant for states within the same $n$ manifold but also
important for states belonging to adjacent $n$ manifolds. 

The above considerations provide an idea how important the
coupling of the CM and electronic motion is for different
values of the parameters $n$(internal energy),$N$(CM energy), 
$B$(field strength). In the case $\frac{\kappa}{\Delta}^{>}_{\sim} 1$
we encounter a strong mixing of the Landau orbitals $\Phi_p^L$ and 
the electronic functions $\psi_q$. The expansion (9) of an eigenfunction $\Psi$ of $\cal{H}$ involves
therefore a large number of wave functions $\lbrace\psi_q\rbrace$ which means
that the typical spatial extension of $\Psi$ with respect to the
electronic coordinates is much larger than that of the individual functions $\psi_q$.
A perturbation theoretical approach with respect to the coupling Hamiltonian
${\cal{H}}_2$ is not appropriate in this case. Instead, as already mentioned above,
an efficient description of the wave function might either be achieved by a
better (intuitive) choice for the functions $\lbrace\psi_{q}\rbrace$ in the expansion (9)
or by pursuing a conceptually different approach (see below).

On the other hand side it is interesting to consider the situation of increasing
internal energy ${\cal{H}}_3$ for fixed field strength. It is well-known
that the fixed nucleus one-electron atom undergoes a classical transition from 
regularity to chaos with increasing excitation energy and the quantized
atom shows the corresponding quantum signatures of chaos \cite{Fri89}.
Perturbation theory with respect to the magnetic interaction terms is only
applicable for sufficiently small field strengths/internal energies which corresponds, in a classical
language, to the regime for which phase space is dominated by regular structures. 
Treating the ion in the intermediate case of mixed regular and chaotic classical phase
space is a difficult task. However for another limiting case, namely the situation
of a completely chaotic phase space
(Coulomb- and diamagnetic interaction are of equal strength) a statistical approach 
to the matrixelements and spectrum of the Hamiltonian ${\cal{H}}_3$ seems appropriate.
Such an approach has been developed in ref.\cite{Lei98} and allows to extract 
important statistical properties for the mixing of the electronic (due to ${\cal{H}}_3$)
and CM (due to ${\cal{H}}_1$) wave functions. The Hamiltonian ${\cal{H}}_3$ is hereby
represented by a random matrix ensemble, i.e. the Gaussian orthogonal ensemble (GOE),
which is the appropriate semiclassical description of a completely chaotic system 
\cite{Boh93}.  The GOE provides the fluctuations of the chaotic levels.
The mean level density (MLD) as a function of energy, field strength and in particular
electronic angular momentum $L_z$ is a key ingredient for the specification of the
random matrix ensemble and can in our case be obtained via the semiclassical Thomas-Fermi
formula \cite{Lei98}. It represents the density of irregular states.
The size of the matrix elements of ${\cal H}_2$ can be determined
from a semiclassical relation between off-diagonal matrix elements of an operator, and the
Fourier transform of its classical autocorrelation function \cite{Fein86}. For more 
details on the concrete appearance of the statistical-semiclassical model for the moving ion
we refer the reader to ref.\cite{Lei98} and report here only on some major results.
Of particular interest are the properties of the eigenvectors which are obtained
through diagonalization of the total matrix Hamiltonian consisting of the parts
${\cal{H}}_1,{\cal{H}}_2$ and ${\cal{H}}_3$. For sufficiently low CM energies 
we encounter an exponential localization of the components of the eigenvectors around
some maximum component. With increasing CM energy the localization length $\lambda$, which reflects
the typical length of the mixing process in the space of the quantum states, increases.
In ref.\cite{Lei98} it was, however, shown that the eigenvectors of this statistical model
possess a finite length $L_c$. At some critical CM energy the localization length becomes
therefore larger than $L_c$, i.e. larger than the size of the system in the space of the
quantum states, and we encounter a crossover from localization to delocalization.
We conclude with the remark that there exists an intriguing analogy of a simplified version of 
the above statistical model with models for transport in disordered
finite-size wires whose localization lengths and related properties are known exactly.

\subsection{Energy transfer processes for rapidly moving\\
atomic ions}

In the present subsection we focus on the situation of a rapidly moving highly excited ion.
The corresponding CM energy is much larger than the initial electronic binding energy.
Since the coupling of the collective and electronic
motion will be large, energy exchange processes between the CM and electronic motion are
extremely relevant and provide some intriguing new phenomena \cite{Schm95a,Schm95b,Mel99a,Mel99b}.
The study of both the classical as well as quantum properties provides additional
insight into our understanding of the quantization on the classical energy flow.

The classical energy exchange equation for the CM energy $E_{cm}$
and for the internal energy $E_{int}=\frac{\mu}{2}\dot{\bf r}^2+V$ reads as follows
\cite{Schm95a,Schm95b}
\begin{equation}
\frac {d}{dt} {E}_{cm} = -\frac {d}{dt} {E}_{int}=
e\alpha \left({\bf{B}}\times {\dot{\bf{R}}}\right)
\dot{\bf{r}}
\end{equation}
This equation shows that a permanent flow of energy from the CM to
the electronic degrees of freedom and vice versa has to be expected.
Let us consider a typical classical trajectory corresponding to the
above-described situation of a rapidly moving highly excited Rydberg atom.
After a transient time of  bound oscillations in the internal motion (energy)
a strong flow of energy from the CM to the internal motion takes
place. The {\it internal} energy is hereby increased above the
threshold for ionization , $E_{th}=0$, and the ion eventually ionizes,
i.e. the electron escapes in the direction parallel to the magnetic field.
Note that the motion of the electron is confined in the direction
perpendicular to the magnetic field.

Figure 3 provides a prototype example for such an ionizing trajectory.
The subfigures 3a and 3b illustrate the time-dependencies of the CM-energy and the
z-component of the internal relative coordinate, respectively.
After the above-mentioned initial phase of oscillations there occurs
at approximately $T=7\times10^{6}$a.u.$(1.7\times10^{-10}s)$
a sudden loss of CM kinetic energy simultaneously accompanied by an increase in the internal energy
which causes the electron to move away from the nucleus in the positive z-direction.
The transferred energy, which is in our case of figure 3 approximately
$6\times 10^{-3}$ a.u.$(0.2eV)$, corresponds to a small fraction of the total initial CM energy
which is for our example about $12.2677$ a.u. $(333.8eV)$.
This self-ionization process is only possible due to the presence of the coupling
term ${\cal H}_2$ in the Hamiltonian (7) which involves both the
internal and CM degrees of freedom. The ionization time for
an individual trajectory depends, apart from its intrinsic dynamics, on
the field strength and in particular on the CM kinetic energy of the ion.

In order to gain an idea of the statistical measure for the ionization process
it is instructive to consider for an ensemble of trajectories the fraction
of ionized orbits as a function of time. The initial internal energy
is chosen to correspond to a completely chaotic phase space of the $He^+-$ion
if the nuclear mass were infinite. In figure 4 we have illustrated
the fraction of ionized orbits as a function of time up to
$T=10^{10}$ a.u. for a series of different CM energies and for
a strong laboratory field strength $B=23.5T$.  For an initial CM energy of $E_{cm}=0.053$ a.u.
which corresponds to an initial CM velocity of $V_{cm}=8.4\times10^3\frac{m}{s}$ about
$70\%$ of the trajectories are ionized within a time of $T=10^{9}$ a.u. ($2.4\times10^{-8}$s)
which is the tenth part of the integration time. In contrast to this we have for
$E_{cm}=0.01$ a.u. only about $30\%$ of ionized orbits within the total
integration time of $T=10^{10}$ a.u. ($2.4\times10^{-7}$s).
The ionization process depends, therefore, very sensitively on the initial
CM kinetic energy of the ion.

The above investigations have shown the existence of the self-ionization process
through energy transfer from the CM to the electronic motion for the classically
moving ion in a magnetic field. For the limit of very highly excited electronic states and
a large CM energies it is expected that the above described behaviour
reflects the true dynamics of the ion. The natural question arises now
how quantization changes or enriches this picture for the typical energies
accessible experimentally. It is well-known that quantization can 
alter the effects observed in classical dynamics (see ref.\cite{Haa91} and references therein).
Very recently an approach which consists of a combination of semiclassical and wave
packet propagation techniques has been developed which is appropriate for
the description of the quantum dynamics of the highly excited and rapidly moving
quantum ion in a magnetic field.  The results of the corresponding investigations \cite{Mel99a}
show that the quantum self-ionization obeys a time scale which is by orders of magnitude
larger than the corresponding classical process. The typical ionization times are for
both the classical as well as quantum processes much smaller than the life times
of the highly-excited Rydberg states. In addition the ionization signal is seemingly affected
by quantum coherence phenomena which yield strong fluctuations for
the time-dependence of the ionization rate. More studies have to be performed in
order to elucidate the dynamics in the semiclassical and deep quantum regime
both theoretically as well as experimentally \cite{Asey96}.

Let us comment on a very recent application \cite{Mel99b} of the effects of
the coupling and mixing of the collective and internal motion in exotic atoms.
It was shown that one can stimulate, by using present-day laboratory
magnetic fields, transitions between the $lm$ sub-levels of fast $\mu He^{+}$ ions
formating in muon catalyzed fusion. This gives a possibility to drive the
population of the $lm$ sub-levels by applying a field of a few Tesla, which affects the
reactivation rate of the fusion process and is especially important to its $K\alpha$
$X-$ray production.

To conclude, we have discussed a number of intriguing phenomena for two-body atoms in 
magnetic fields which have their origin in the nonseparability
of the CM and internal motion causing an interaction of the collective
and internal motion. It has to be expected that this interaction will induce
even more phenomena for multi-electron atoms or molecules. For the
case of heteronuclear molecules the relevant coupling Hamiltonian contains
both the vibrational and rotational degrees of freedom and new vibrational
or rotational structures as well as dynamical processes might therefore
arise once the coupling becomes significant i.e. comparable to the
spacing of the vibrational or rotational energy levels.
\vspace*{1.0cm}

%{\large \bf Acknowledgments}
%
%One of the authors (P.S.) acknowledges financial support by the National
%Science Foundation through a grant (P.S.) for the Institute for Theoretical
%Atomic and Molecular Physics at Harvard University and Smithsonian Astrophysical Observatory.
%\vspace*{0.5cm}

\section*{Figure captions}

{\bf{Figure 1:}}
A two-dimensional intersection of the potential $\cal{V}$ in the plane
perpendicular of the magnetic field.
\vspace{0.5cm}

{\bf{Figure 2:}}
Distribution over energy in the outer well after the preparation is
completed. In the calculation a magnetic field strength $B=14 T$ 
was used. The ionization threshold is shown by a vertical dashed line.
The inset shows the spreading of the trajectories over the saddle point
into the outer well during the intermediate time step of a constant
electric field $E_c$.
\vspace{0.5cm}

{\bf{Figure 3:}}
(a) The CM energy as function of time. (b) The z-component of the internal
relative coordinate as a function of time.  The total and initial internal
energy of the ionizing trajectory are $E=333.68$eV and $E_{int}=-8.16\times10^{-2}$eV,
respectively. The field strength is $B=23.5$T.
\vspace{0.5cm}

{\bf{Figure 4:}}
The ionized fraction for an ensemble of $250$ trajectories as a function
of time.From top to bottom the CM-energies belonging to the  ionization
curves are $E_{CM}=1.45,0.63,0.47,0.34,0.27 eV$, respectively. The initial
internal energy is always $E_{int}=-9.25\times10^{-3}$eV. The field
strength is  $B=23.5$T.

\newpage

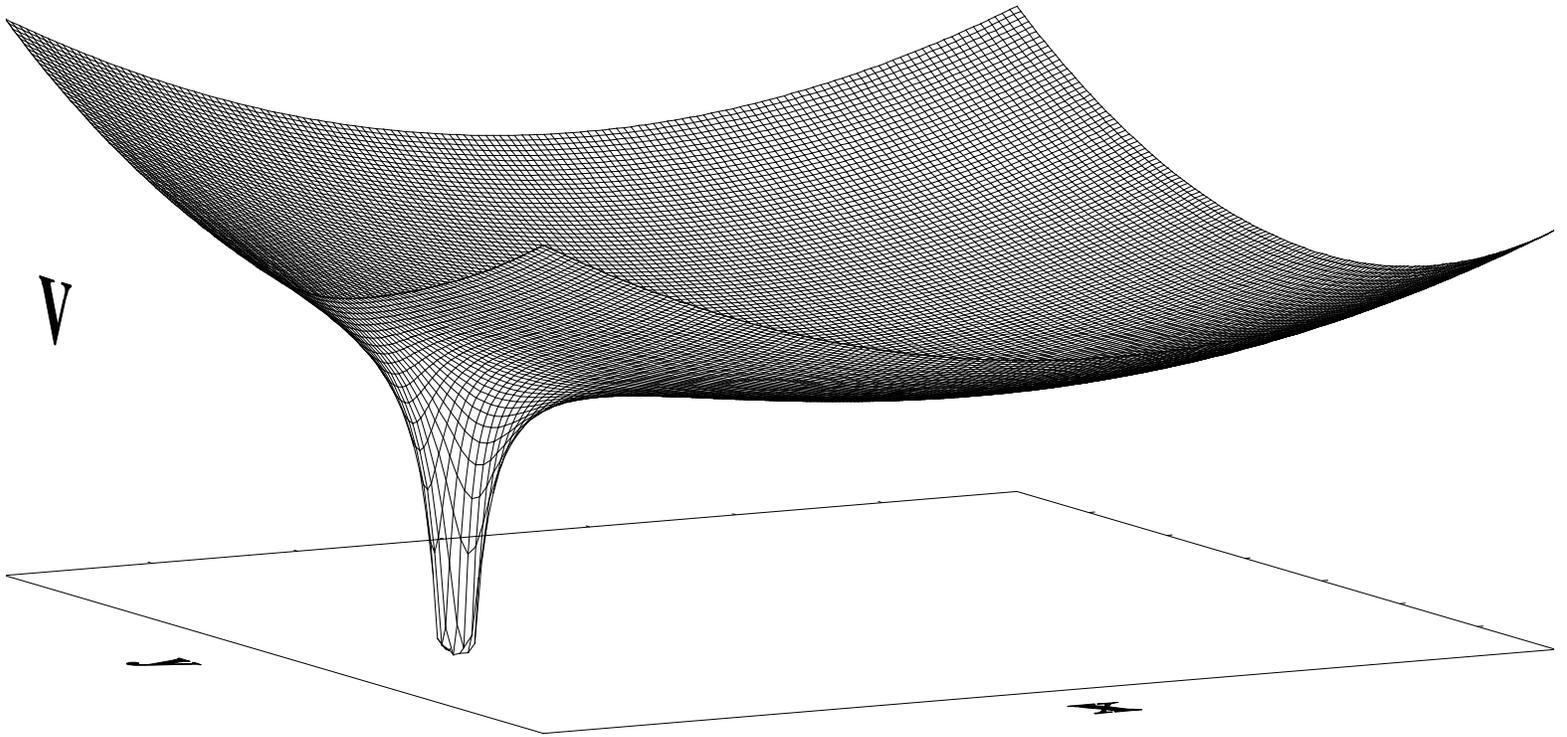
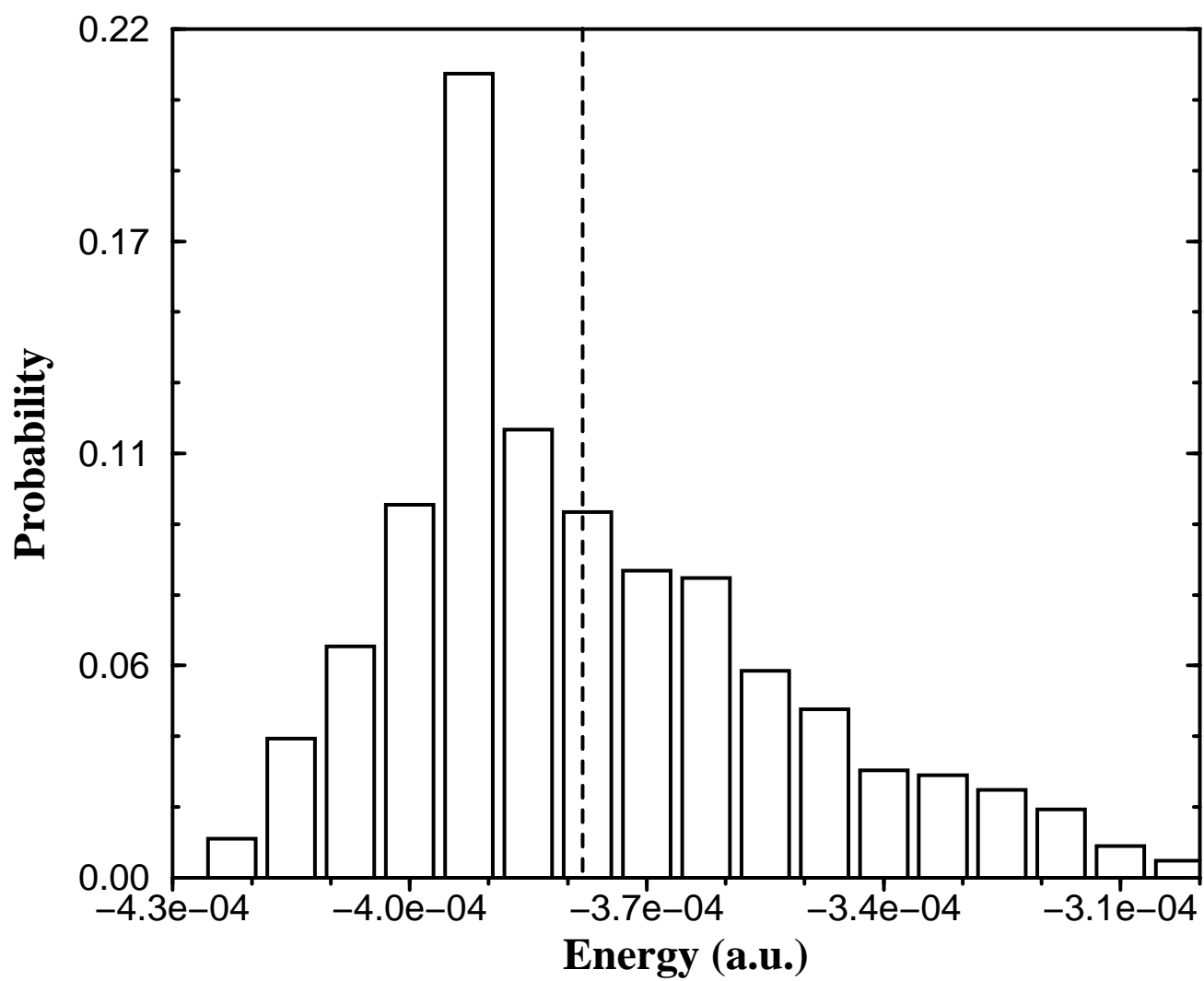
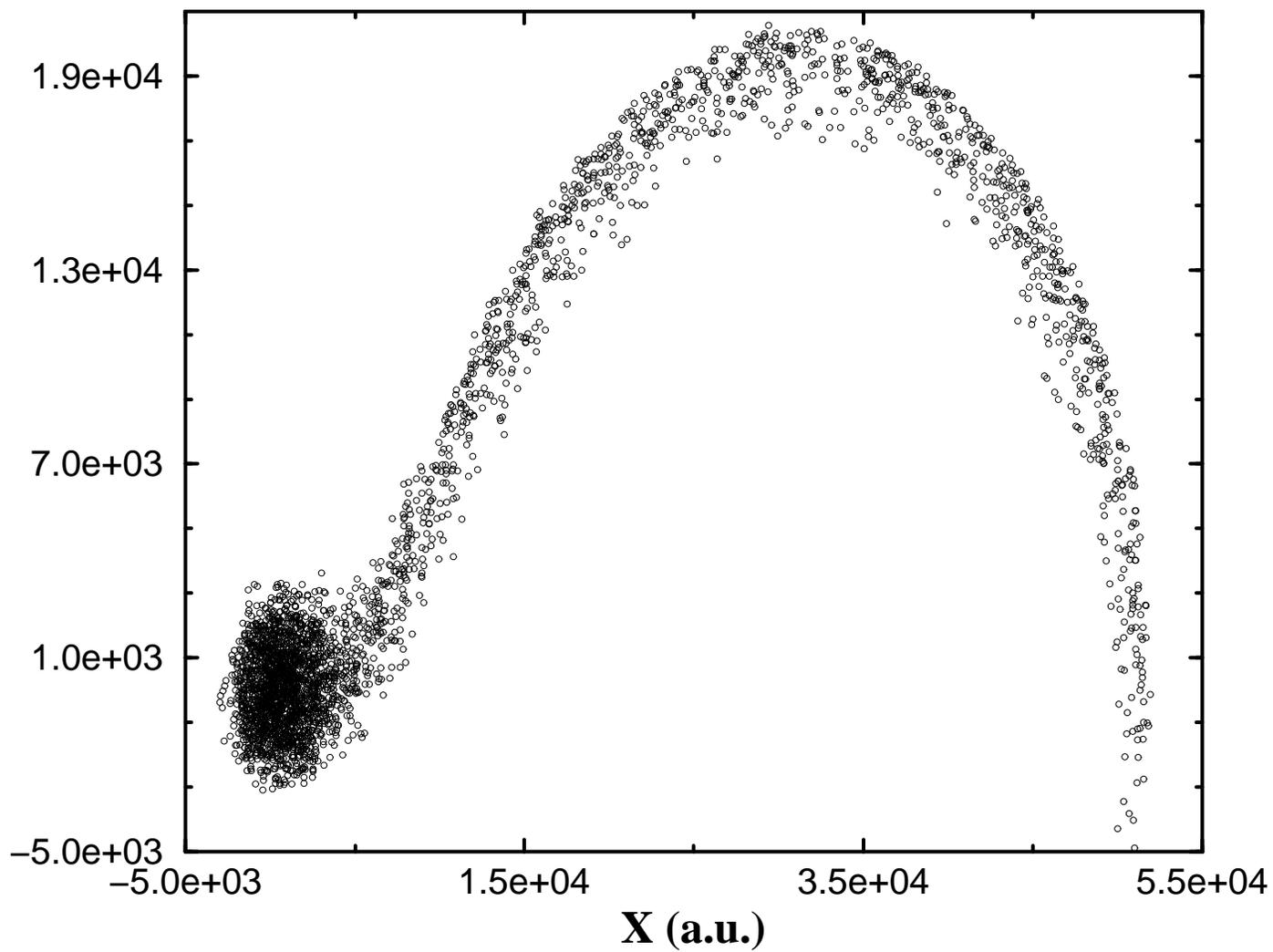
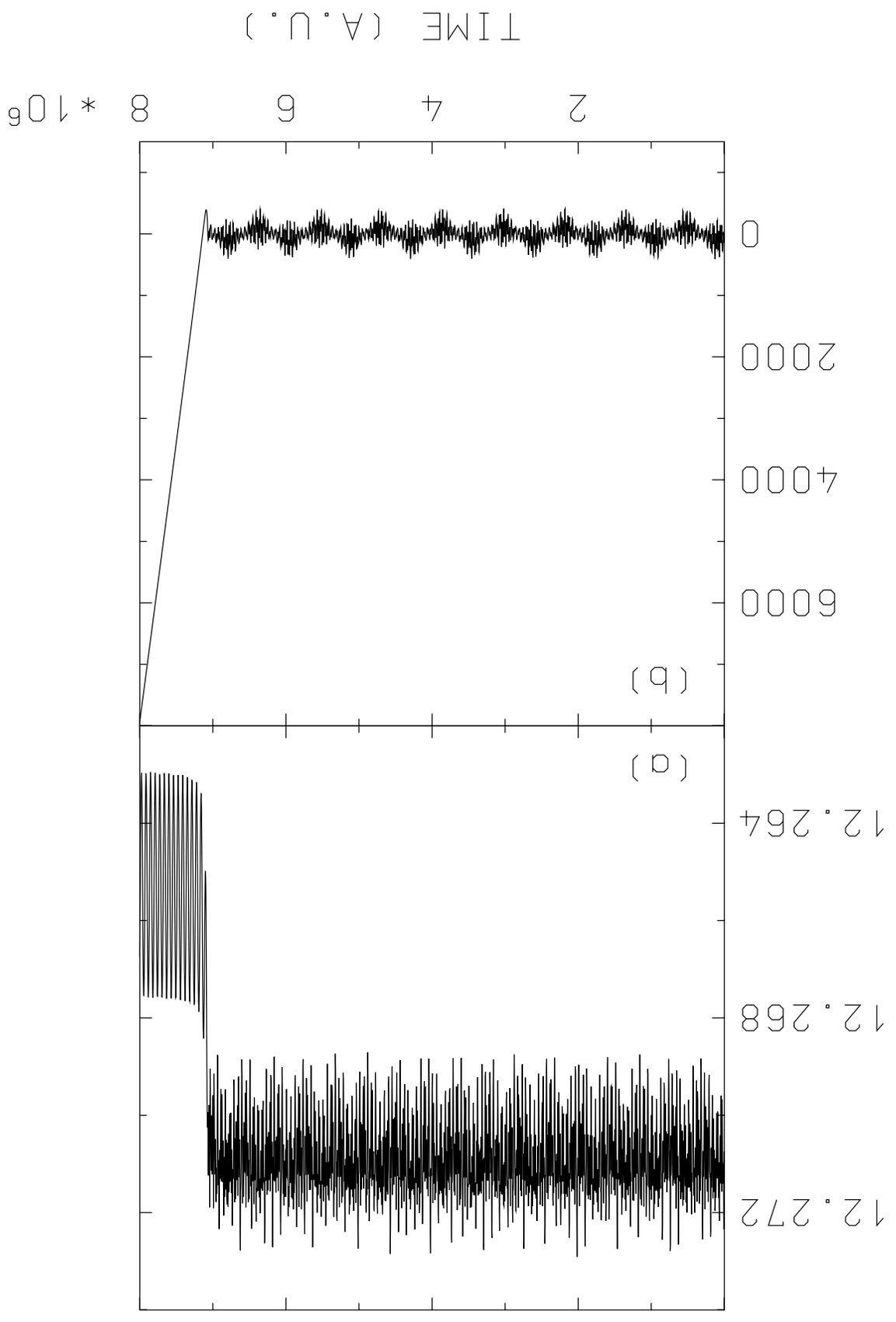
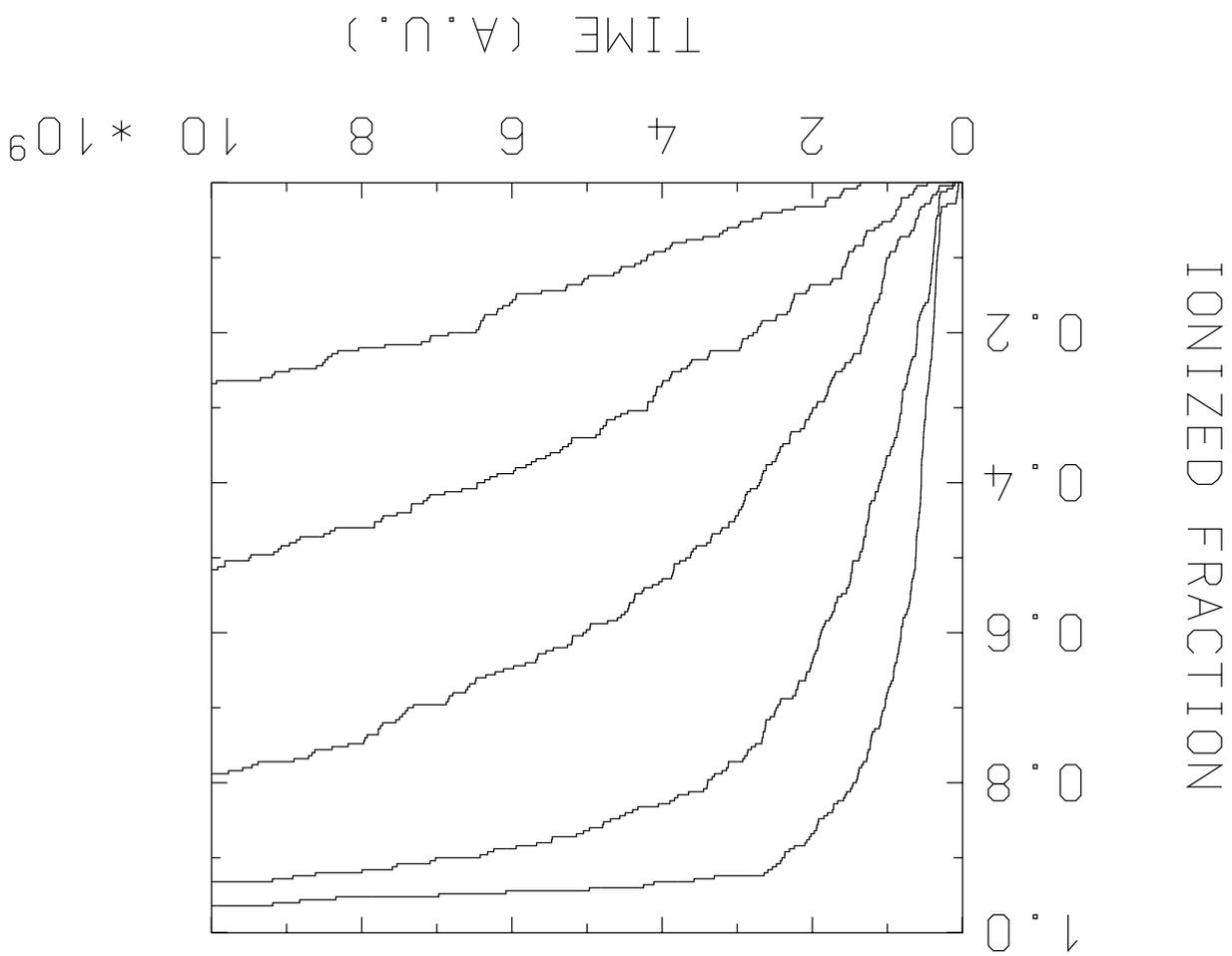
\end{document}